\documentclass[prl,aps,floatfix,twocolumn,amsmath,amssymb,superscriptaddress,showpacs]{revtex4-1}
\usepackage{graphicx}
\usepackage{bm}
\usepackage{amsmath}
\usepackage{dcolumn}
\usepackage{dsfont}
\usepackage{amssymb}
\usepackage{color}
\usepackage{epstopdf}
\usepackage{epsfig}
\usepackage{mathrsfs}
\usepackage{lineno}
\usepackage{setspace}
\usepackage{physics}

\usepackage[colorlinks=true, pdfstartview=FitV, linkcolor=blue, citecolor=blue, urlcolor=blue]{hyperref}
\setcitestyle{super,compress}

\newcommand{\yxadd}[1]{\textcolor{black}{#1}}

\begin{document}
\title{Superfast hole spin qubits enabled by uniaxial strain-boosted spin-orbit coupling}

\author{Yi-Xu Wang$^{\text{1, 2, }*}$,Yang Liu}
\email{These authors contributed equally to this work}
\affiliation{State Key Laboratory of Semiconductor Physics and Chip Technologies, Institute of Semiconductors, Chinese Academy of Sciences, Beijing 100083, China}
\affiliation{Center of Materials Science and Optoelectronics Engineering, University of Chinese Academy of Sciences, Beijing 100049, China}

\author{Shan Guan}
\email{shan\_guan@semi.ac.cn}
\affiliation{State Key Laboratory of Semiconductor Physics and Chip Technologies, Institute of Semiconductors, Chinese Academy of Sciences, Beijing 100083, China}
\author{Jun-Wei Luo}
\email{jwluo@semi.ac.cn}
\affiliation{State Key Laboratory of Semiconductor Physics and Chip Technologies, Institute of Semiconductors, Chinese Academy of Sciences, Beijing 100083, China}
\affiliation{Center of Materials Science and Optoelectronics Engineering, University of Chinese Academy of Sciences, Beijing 100049, China}
\author{Shu-Shen Li}
\affiliation{State Key Laboratory of Semiconductor Physics and Chip Technologies, Institute of Semiconductors, Chinese Academy of Sciences, Beijing 100083, China}
\affiliation{Center of Materials Science and Optoelectronics Engineering, University of Chinese Academy of Sciences, Beijing 100049, China}

\begin{abstract}
 Two-dimensional (2D) electron/hole gases confined in semiconductor heterostructures suffer from weak Rashba spin-orbit coupling (SOC) for manipulating spin degreee of freedom via an electric rather than a magnetic field. Here, we show that complementary metal-oxide-semiconductor technology-accessible strain could substantially enhance the linear Rashba SOC of the top hole subband in Ge/SiGe quantum wells (QWs) to a level comparable to that of 2D Rashba materials through enhancing the mixture of the light-hole and heavy-hole bands. We further show that strongly enhanced Rashba SOC boosts the Rabi frequency of hole spin qubits confined in Ge/SiGe QWs by two orders of magnitude to an unprecedented 40 GHz, more than one order of magnitude faster than other qubit platforms. We also demonstrate that the hole spin rotation with Rabi frequency $\gtrsim 25$~GHz enters a new regime being immune to gate control-induced electric noise, opening a new avenue to simultaneously improve the gate speed and gate fidelity. Our findings provide a new routine to substantially enhance the Rashba SOC in 2D semiconductor hole gases to a level that is great for spintronic applications.
\end{abstract}

\maketitle

\yxadd{Spintronics promise to revolutionize electronics and computing by making explicit use of the spin of electrons in addition to their electrical charge by creating new spin-based electronic devices or enhancing the functionality of devices by offering higher speed of data processing, greater data storage density, and lower energy consumption~\cite{Zutic_RMP2004}.
A key requirement of spintronics is spin manipulation via electric rather than magnetic fields, which could be achieved by the electrically tunable Rashba spin-orbit coupling (SOC) that entangles the spin and orbital degrees of freedom.
It has inspired a wide range of predictions, discoveries, and concepts, including spin transistors, spin-orbit qubits, the spin Hall effect, the quantum spin Hall effect, topological insulators, and Majorana fermions~\cite{Datta1990, Schliemann2003, Maurand2016, Watzinger2018, Sinova2004, DiXiao2010, Bernevig2005, Bernevig2006, Awschalom2013, Sarma2010, Mourik2012, Manchon2015, King_PRL2011}.
However, despite decades of research, demonstrating such control even in the prototypical system of two-dimensional electron gases (2DEGs) via Rashba SOC remains difficult to implement~\cite{Manchon2015}, because weak Rashba SOC in most semiconductors necessitates low temperatures for device operation and long channel lengths with ultrahigh purity material to avoid spin-flip scattering events. It has thus caused to look for strong Rashba SOC in metal surfaces~\cite{LaShell_PRL1996, Koroteev_PRL2004, Ast_PRL2007}, in ultrathin metal films~\cite{frantzeskakis2008, he2008} and in 2D materials~\cite{bordoloi2024}.}

{Although linear Rashba SOC was considered to be absent in [001]-oriented semiconductor quantum wells (QWs), it has been recently demonstrated that their local interface with a lower symmetry could introduce a finite linear Rashba SOC in their top hole subband~\cite{Xiong_PRB2021} to enable the fast qubit control of planer Ge hole spin qubits confined in planar Ge quantum dots (QDs) hosted in [001]-oriented Ge/SiGe QWs via electric dipole spin resonance (EDSR)~\cite{EDSR2007, Hendrickx2020,  LiuyangPRB2022}. Planer Ge hole spin qubits have further been demonstrated with single-qubit gate fidelity beyond 99.99\% (far exceeding the threshold 99\% required by surface code correction) at an operation speed above 50~MHz~\cite{lawrie2023}, two-qubit gate fidelity of 99.3\%~\cite{Wang2024}, charge manipulation of a 16 QD array~\cite{Borsoi2024}, and coherent control of a 10 qubit array~\cite{Wang2024}. Additionally, the two-dimensional hole gases (2DHGs) in Ge QWs have been demonstrated to have long spin relaxation time~\cite{giorgioni2016} and exceptionally high mobilities exceeding $10^6~\mathrm{cm^2\,V^{-1}\,s^{-1}}$~\cite{myronov2023a}. Combined with the compatibility with the modern microelectronic industry, these progresses have demosntrated 2DHGs in Ge QWs being ideal playgrounds to tailor functional properties through SOC for spintronics.}

\begin{figure*}[t]
	\centering
	\includegraphics{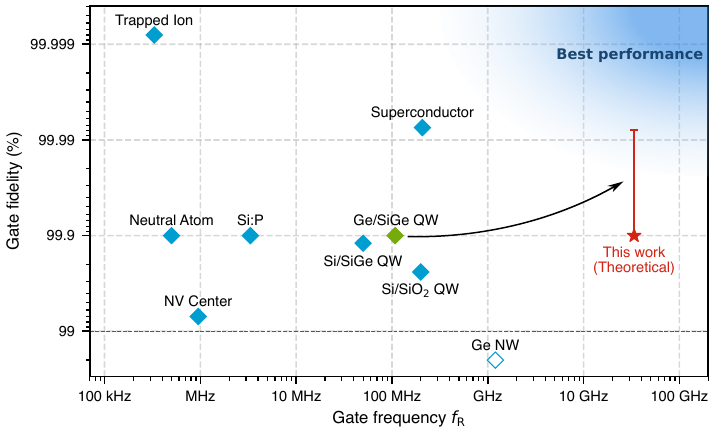}
	\caption{\textbf{Comparison of single-qubit gate fidelity and operation speed across scalable qubit platforms (2020 - 2026).}  Reported results are shown for superconducting qubits~\cite{Anferov2024, li2023}, trapped ions~\cite{loschnauer2025}, neutral atoms~\cite{bluvstein2026}, nitrogen-vacancy (NV) centers in diamond~\cite{Vallabhapurapu2023}, and semiconductor spin qubits~\cite{camenzind2022, Ha2022, Nakajima2020}. Hole spin qubits hosted by Ge Nanowires (NW)~\cite{Liu2023} exhibit sizable operation speed, although the gate fidelity was not reported and is therefore assumed to lie below the surface-code threshold of 99\%. This work presented superfast hole spin qubit (marked by the red star) is plotted using the operation speed at the ``sweet spot'' (uniaxial strain of $\epsilon_{\text{u}}=0.2\%$ along the [110] direction, biaxial strain of $\epsilon_{\text{b}}=-0.1\%$, vertical electric field of $E_z=152$ kV cm$^{-1}$) with the gate fidelity assumed to be in the possible range. The comparison highlights the general trend in performance among leading qubit platforms, but it does not represent an exhaustive survey or an assessment of the quality of individual works.}
	\label{fig:fig1} 
\end{figure*}

Despite the rapid advance of Ge hole spin qubits, it still lags behind the widely adopted superconducting qubits~\cite{kjaergaard2020} because of the relatively low operation speed of Ge hole spin qubits limited by the weak linear Rashba SOC in Ge QWs. The single-qubit operation speed of Ge hole spin qubits is primarily limited by the strength of the direct linear Rashba SOC, which is proportional to the strength of heavy-hole-light-hole (HH-LH) mixing that is weak in the [001]-oriented Ge/SiGe QWs because it is forbidden by the global QW symmetry and is allowed by the local interface alone. In this work, we propose to apply the [110] uniaxial strain, a standard complementary metal-oxide-semiconductor (CMOS) technology process for improving hole mobility, to lower the symmetry of the [001]-oriented Ge/SiGe QWs to enhance the linear Rashba SOC and thus boost the Rabi frequency of planer Ge hole spin qubits. By performing the atomistic semi-empirical pseudopotential method (SEPM) calculations, we show that a CMOS technology accessible uniaxial strain can singnificantly enhance the linear SOC strength of the top hole subband in [001]-oriented Ge/SiGe QWs to a level comparable to that of 2D Rashba materials~\cite{bordoloi2024}. Based on the model Hamiltonian analysis, we unambiguously show that this enhancement results from uniaxial strain-induced HH-LH mixing and remains hindered by the residual -0.61\% biaxial strain presented in the Ge well. By reducing the residual biaxial strain to -0.1\%, we show that the uniaxial strain could further boost the linear SOC. The strong SOC accelerates the Rabi frequency of the planar Ge hole qubits, surpassing all types of known qubits~\cite{Anferov2024, bluvstein2022, Vallabhapurapu2023, Ha2022, Liu2023}, into the 10~GHz regime, enabling planar Ge hole qubits operating at completely new timescales. We also show the emergence of a maximum in the strain-enhanced linear SOC strength as a function of the applied electric field, offering a ``sweet spot'' designed to suppress the magnified decoherence induced by a stronger SOC, which would otherwise overwhelm the benefits of fast operation. \yxadd{Furthermore, we explore the previously unexplored spin dynamics in the strong-SOC limit and identify a new operational regime for simultaneously high-fidelity and high-speed single qubit operation.}

The performance and scalability of a quantum computer are limited by the number of operations performable within the qubit coherence lifetime~\cite{howard2023}, which renders accelerating the operation speed to a primary focus in the development of qubit hardware, leading to intensive exploration across device technologies, such as superconductors~\cite{rower2024, howard2023}, trapped ions or neutral atoms~\cite{saner2023, gale2020}, and semiconductors~\cite{Wang_NC2022, Liu2023}.  Fig.~\ref{fig:fig1} summarizes the fidelity and speed of the state-of-the-art qubit devices for these typical types of qubit platforms. It shows that high-fidelity superconducting qubits operations at 0.2~GHz, surpassing other qubit platforms and thus positioning superconducting quantum computers as the most advanced and widely adopted qubit platform~\cite{kjaergaard2020}.
In contrast, the speed of the planar Ge hole spin qubits is still too slow considering the low fidelity.
Furthermore, compared with the two-qubit gate for Ge hole spin qubits, the single-qubit gate speed has become the main bottleneck to incease the operation speed, which is evidenced by the sharp difference between CZ ($9~\text{ns}$) and CX ($100~\text{ns}$) gates with the latter being limited by slow single-qubit gates ($\gtrsim 30~\text{ns}$)~\cite{Hendrickx2021}.
Therefore, speeding up the gate operations and improving the fidelity of qubits are central tasks. Here, we explore approches to enhance linear Rashba SOC of the top hole subband in Ge/SiGe QWs in order to boost the EDSR rabi frequency of planer Ge hole qubits by performing the atomistic SEPM.

{\bf Uniaxial strain remarkably enhances the direct Rashba SOC}

\begin{figure*}
  \centering
  \includegraphics{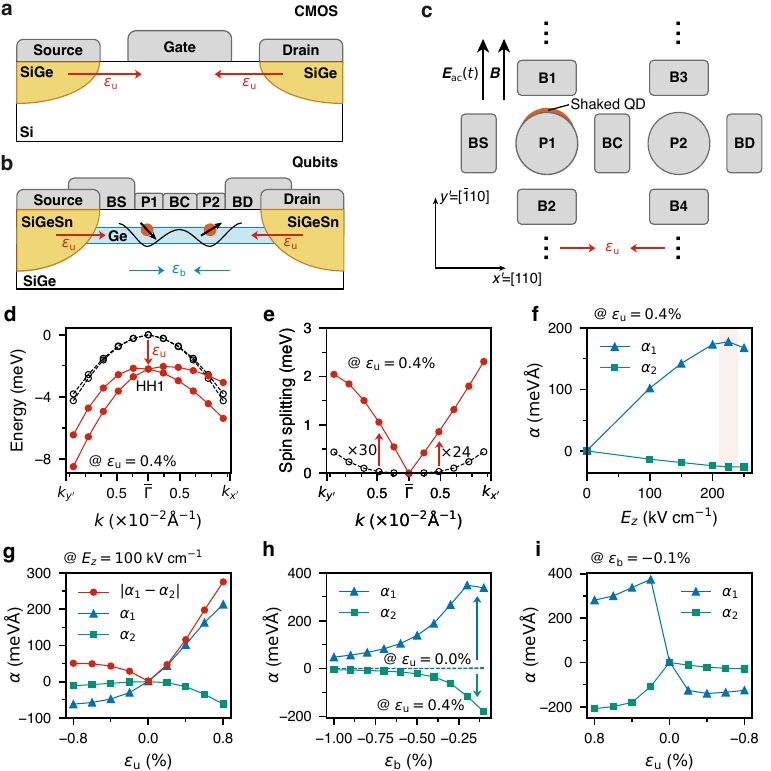}
  \caption{\textbf{CMOS-compatible strain engineering substantially enhances linear SOC.}
    (\textbf{a}) Strain-engineering in CMOS technology to enhance hole mobility by embedding larger-lattice SiGe alloy in the source and drain regions, generating a compressive [110] uniaxial strain in the Si channel.
    (\textbf{b}) Analogous CMOS-compatible strain-engineering approach to apply uniaxial strain to Ge/SiGe QW hosting spin qubit arrays by embedding alloy in the source and drain regions. BS, BD, and BC denote the source, drain, and center barrier gates; P1 and P2 denote the plunger gates~\cite{Hendrickx2020}.
    (\textbf{c}) Schematic of linear-SOC-mediated EDSR in a planar QD array. The uniaxial strain and magnetic field are applied along the $x'$=[110] and $y'$=[$\overline{1}1$0] directions, respectively. Top and bottom gates (140 nm apart) generate an ac electric field $E_{y'}(t)=E_\text{ac}\cos\omega_\text{ac} t$.
    (\textbf{d}) Energy dispersion of the top hole subband (HH1) before (black dashed) and after (red solid) applying uniaxial strain of $\epsilon_{\text{u}} = 0.4\%$ along the [110] direction under a vertical bias electric field $E_z$ = 100 kV cm$^{-1}$.
    (\textbf{e}) Corresponding spin splitting.
    (\textbf{f}) Under a fixed $\epsilon_{\text{u}}$=0.4\%, linear SOC parameters $\alpha_{\text{1}}$ and $\alpha_{\text{2}}$ as functions of $E_{z}$, highlighting a ``sweet spot''.
    (\textbf{g-i}) $\alpha_{\text{1}}$ and $\alpha_{\text{2}}$ in the Ge QW (\textbf{g}) as functions of $\epsilon_{\text{u}}$ at $\epsilon_{\text{b}} = -0.61\%$, (\textbf{h}) as functions of $\epsilon_{\text{b}}$ at $\epsilon_{\text{u}} = 0.4\%$ and $0.0\%$, (\textbf{i}) as functions of $\epsilon_{\text{u}}$ at $\epsilon_{\text{b}} = -0.1\%$.}
  \label{fig:fig2}
\end{figure*}

First of all, the atomistic SEPM is used to calculate the energy dispersion of the top hole subband for a gate-biased Ge/SiGe QW, which has recently been used to host the planar hole spin qubits for the experimental demonstration of 108~MHz Rabi frequency~\cite{Hendrickx2021}.
\yxadd{The SEPM is employed here to account for all atomistic details and potential symmetry-breaking effects, unlike recent studies relying on model Hamiltonians~\cite{venitucci2018, terrazos2021, martinez2022, Abadillo-Uriel2023, Rodriguez-Mena2023, mauro2025, mauro2025a}, where the underlying approximations may affect the results and obscure key physics~\cite{Zunger2002, Luo_PRL2009, LuoJW_PRL2010, Luo2010, Luo2011, Luo_PRB2015, Luo2017, Xiong_PRB2021}.}
Specifically, the Ge/SiGe QW consists of a 120-monolayer (ML) thick Ge well embedded in the $\mathrm{Si_{0.2}Ge_{0.8}}$ alloy barrier and is under an electric field of $E_z$ = 100~kV\,cm$^{-1}$ normal to the substrate. After relaxing the atom positions, we find that this 120-ML thick Ge well is in a compressive biaxial strain of $\epsilon_{\text{b}}=-0.61\%$, consistent with experimentally observed $\epsilon_{\text{b}}=-0.63\%$~\cite{Corley-Wiciak2023}. The top hole subband derives mainly from the bulk HH band and is thus denoted as HH1. The atomistic SEPM calculations show that the strain-dependent effective mass of the HH1 subband is $0.124$~$m_0$ ($m_0$ is the bare electron mass), which is also in good agreement with the experimentally measured value of $0.105$~$m_0$~\cite{Laroche2016}. 

Fig.~\ref{fig:fig2}d shows that HH1 possesses two spin branches with an increased energy separation as the in-plane wavevector $\bf {k_{||}}$ goes away from the $\overline{\Gamma}$ point, which is referred to as momentum-dependent spin splitting. Fig.~\ref{fig:fig2}e shows the corresponding spin splitting of HH1 as a function of $k$ along [110] and [$\overline{1}1$0] directions, respectively. One can see that the spin splitting is predominantly $k$-linear, consistent with the predictions of the newly discovered direct Rashba SOC~\cite{Kloeffel2011, Luo2011}, which originates from the interplay between HH-LH mixing and the direct dipolar coupling of an external electric field to the hole subbands in low-dimensional hole gases~\cite{Kloeffel2011, Luo2011, Luo2017, Xiong_PRB2021}.  The newly discovered linear direct Rashba SOC is of particular importance to achieve fast all-electric spin manipulation via EDSR since it potentially couples the spin to the momentum of the hole by creating an effective periodic magnetic field when the dot is shaken as a whole under an ac electric field~\cite{EDSR2007, LiuyangPRB2022}. However, this direct Rashba SOC is relatively weak in the CMOS-compatible [001]-oriented Ge/SiGe QWs because its strength is proportional to the strength of HH-LH mixing~\cite{Luo2011, Luo2017}, which is induced by the local $C_{2v}$ interfaces alone in [001]-oriented Ge/SiGe QWs. This is because the global symmetry of the QW forbids the HH-LH mixing due to both HH and LH states belonging to different representations in the higher symmetry point group of the QWs, but belonging to the same $\Gamma_5$ irreducible representation under $C_{2v}$ point group of the local interface symmetry~\cite{Xiong2022}. Fig.~\ref{fig:fig2}f shows that the linear Rashba parameter is predicted to be 1.23~meV\,\AA~for the HH1 in the 120 ML thick Ge QW. We have demonstrated that just this weak $k$-linear Rashba SOC enables a fast EDSR Rabi frequency of 108 MHz achieved in the recent experiment for hole spin qubits hosted by the 120 ML thick Ge/Si$_{0.2}$Ge$_{0.8}$ QW~\cite{Xiong_PRB2021,LiuyangPRB2022, Hendrickx2021}. 

To obtain a stronger Rashba SOC for a faster Rabi frequency, we can lower the QW symmetry to introduce strong HH-LH mixing. For instance, we have shown that the direct linear Rashba SOC is 50 times stronger in [110]-oriented Ge/SiGe QWs than that in [001]-oriented counterparts because the HH-LH mixing is allowed by the former QW global symmetry~\cite{Xiong2022}. However, the [110]-oriented Ge/SiGe QWs are incompatible with the mature CMOS technology, which is exclusively based on the (001) Si wafer. To speed up the planar Ge hole qubits in a way compatible with CMOS technology, here, we propose to utilize the uniaxial strain to lower the QW symmetry to introduce HH-LH mixing and hence enhance the linear Rashba SOC in the [001]-oriented Ge/SiGe QWs. The compressive uniaxial strain in the Si channel along [110] direction has been introduced to enhance hole mobility in pMOSFET since the 90-nm CMOS technology node in 2003 by embedding SiGe alloy (larger lattice constant) in adjacent source-drain (S/D) regions (as shown in Fig.~\ref{fig:fig2}a)~\cite{bedell2014} \yxadd{and has been utilized in cryogenic CMOS devices~\cite{casse2009, li2011}.} The amplitude of the compressive uniaxial strain exerted on the Si channel depends linearly on the Ge concentration in the S/D regions, achieving at least $1 \%$ uniaxial strain in several ${\rm \mu m^2}$ area~\cite{bedell2014, sawano2008, Lee2004}. Similarly, we can embed SiGe (smaller lattice constant) or GeSn (larger lattice constant) alloy into the S/D regions to create a tensile or compressive [110] uniaxial strain in the Ge well region that hosts the hole spin qubit array, as shown in Fig.~\ref{fig:fig2}b. \yxadd{Direct experimental validation of 0.4\% uniaxial strain in 5~$\mu$m wide Ge/SiGe striplines~\cite{sawano2016} and the successful integration of GeSn stressors in Ge CMOS architectures~\cite{takeuchi2011} confirm that the required strain parameters are achievable within high-quality, mature fabrication environments.}
\yxadd{Moreover, the proposed SOC enhancement is independent of the specific strain-inducing architecture, as the underlying physical mechanism remains valid across various structural implementations~\cite{woods2024, mauro2025a}.}
In contrast to other proposals, such as squeezed qubit~\cite{Bosco2021} and inhomogeneous strain~\cite{Abadillo-Uriel2023}, our proposal features high compatibility with the CMOS technology.

Fig.~\ref{fig:fig2}d shows that an uniaxial strain $\epsilon_{\text{u}} =0.4 \%$ applied along the [110]-direction indeed remarkably enlarge the energy separation of two HH1 spin branches with corresponding $k$-dependent spin splitting enhanced by a factor of about 30 and 24 for the in-plane wavevector along [$\overline{1}$10] and [110] directions, respectively, as shown in Fig.~\ref{fig:fig2}e. \yxadd{Here, uniaxial strain $\epsilon_{\text{u}}$ is defined as $\epsilon_{\text{u}}=(l-l_0)/l_0\times 100\%$, where $l_0$ and $l$ denote the cell lengths along the [110] direction before and after the application of uniaxial strain, respectively. The relaxation results based on the VFF method shows $\epsilon_{\text{u}}\approx 2\epsilon_{xy}$ (See Supplementary Note 2).} This anisotropy has never been reported in 2D heterostructures since the conventional Rashba and Dresselhaus SOC are in-plane isotropic. To account for the in-plane anisotropy in the spin splitting, we employ the invariant theory to derive the 2$\times$2 effective SOC Hamiltonian $H_{\text{soc}}$ for HH1.  In farsightedness of the effect of alloy randomness,  the Ge/Si$_{0.2}$Ge$_{0.8}$ QW under $\epsilon_{\text{u}}$ and $E_z$ belongs to a $C_{2v}$ point group, and  its effective SOC Hamiltonian $H_{\text{soc}}$ thus reads to the lowest order as~\cite{Rodriguez-Mena2023}
\begin{equation}\label{Heff}
H_{\text{soc}}=\alpha_\text{1}(k_{x}\sigma_{x}+k_{y}\sigma_{y})+\alpha_\text{2}(k_{x}\sigma_{y}+k_{y}\sigma_{x}),
\end{equation}
where $\sigma_{x,y}$ are the Pauli matrices, $\alpha_{\text{1}}$ and $\alpha_{\text{2}}$ are linear parameters of two $E_z$-induced direct Rashba SOC terms: the first term is referred to as Dresselhaus-like term and the second one is referred to as Rashba-like term. Fig.~\ref{fig:fig2}e shows that the combination of these two terms makes spin splitting anisotropy with $2(\alpha_\text{1}+\alpha_\text{2})|k|$ along the [110] direction and $2(\alpha_\text{1}-\alpha_\text{2})|k|$ along the [$\overline{1}$10] direction, respectively.  We can assess these two terms by evaluating the $\alpha_\text{1}$ and $\alpha_\text{2}$ coefficients by fitting atomistic SEPM predicted spin splitting to equation~(\ref{Heff}).  

In the absence of uniaxial strain, we obtain $\alpha_{\text{1}}=1.23$~meV\,{\AA} and $\alpha_{\text{2}}=0.01$~meV\,{\AA} for the [001]-oriented 120-ML thick Ge QW under $E_z=100$~kV\,cm$^{-1}$. Fig.~\ref{fig:fig2}g shows that the tensile uniaxial strain rapidly increases both $\alpha_{\text{1}}$ and $\alpha_{\text{2}}$ by more than two orders of magnitude, manifesting as they raise to $\alpha_{\text{1}}=102.28$~meV\,{\AA} and $\alpha_{\text{2}}=-6.70$~meV\,{\AA} at $\epsilon_{\text{u}}=0.4\%$. In contrast, the compressive uniaxial strain is less efficient as the magnitude of $\alpha_{\text{1}}$ and $\alpha_{\text{2}}$ raise only to $-47.7$ and $-4.45$~meV\,{\AA}, respectively, at $\epsilon_{\text{u}}=-0.4\%$. Fig.~\ref{fig:fig2}f shows the predicted $\alpha_{\text{1}}$ and $\alpha_{\text{2}}$ as a function of $E_z$ for the QW under a fixed uniaxial strain of $\epsilon_{\text{u}}=0.4\%$. It is ready to observe that $\alpha_{\text{1}}$ tends to a steady value of 178 meV\AA~at $E_z=225$~kV\,cm$^{-1}$, indicating the emergence of the maximum against the applied electric field, which is a fingerprint of strong direct Rashba SOC~\cite{Luo2017}.
 
{\bf Reducing biaxial strain further enhances SOC}

\begin{figure}[t]
  \centering
  \includegraphics[width=\linewidth]{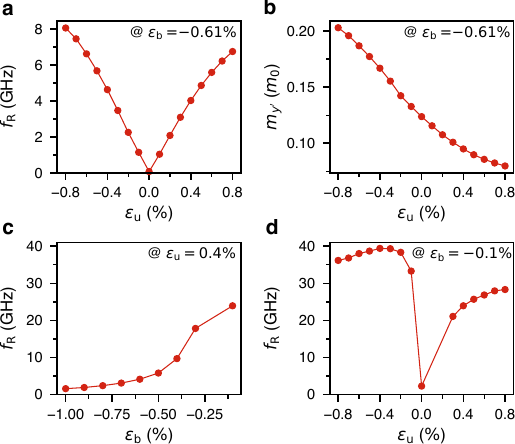}
  \caption{\textbf{Strain engineering significantly boosts gate speed in planar Ge hole spin qubits.} (\textbf{a}) Estimated EDSR Rabi frequency $f_{\text{R}}$ based on obtained $\alpha_{\text{1}}$ and  $\alpha_{\text{2}}$ as a function of uniaxial strain for the strain configurations adopted in Fig.~2\textbf{g}. (\textbf{b}) The corresponding effective mass $m_{y'}$ as a function of uniaxial strain. (\textbf{c}) Estimated EDSR Rabi frequency $f_{\text{R}}$ as a function of biaxial strain for the strain configurations adopted in Fig.~2\textbf{h}. (\textbf{d}) Estimated Rabi frequency $f_{\text{R}}$ as a function of uniaxial strain for the strain configurations adopted in Fig.~2\textbf{i}.}
  \label{fig:biax}
\end{figure}

Based on the model analysis (see the Supplementary Note 3, equations~(S9) and (S13)), we find that reducing the residual biaxial strain $\epsilon_{\text{b}}$ further enhances the linear SOC. Fig.~\ref{fig:fig2}h shows the predicted $\alpha_{\text{1}}$ and $\alpha_{\text{2}}$ with varying the compressive biaxial strain $\epsilon_{\text{b}}$ from -1.0\% to -0.1\% but fixing the uniaxial strain to $\epsilon_{\text{u}}=0.4\%$. One can see that the magnitude of both $\alpha_{\text{1}}$ and  $\alpha_{\text{2}}$ increase remarkably by reducing the compressive biaxial strain $\epsilon_{\text{b}}$. Note that $\alpha_{\text{1}}$ tends to saturation when the magnitude of the residual compressive biaxial strain reduces to less than 0.2\% due to the approach of $\left|B_n\right|$ to $\left|D\right|$ under smaller $\epsilon_{\text{b}}$ (equation~(S9)). To explore even stronger linear SOC, we also examine both $\alpha_{\text{1}}$ and  $\alpha_{\text{2}}$ by fixing the biaxial strain to a small level $\epsilon_{\text{b}}=-0.1\%$ but varying the uniaxial strain $\epsilon_{\text{u}}$. Fig.~\ref{fig:fig2}i shows that $\alpha_{\text{1}}$ has a sharp change around zero uniaxial strain but quickly tends to a saturation value of 375~meV\,{\AA} when $\epsilon_{\text{u}}>0.2\%$ and to a saturation value of 140~meV\,{\AA} when $\epsilon_{\text{u}}<-0.2\%$. Such tuning of biaxial strain can be achieved in experiments by changing the alloying components in the buffer layer without introducing new structures~\cite{Lodari2022, costa2025}.
Notably, the linear Rashba SOC obtained here is comparable in magnitude to that of promising 2D Rashba materials, whose linear Rashba parameters range from 12~meV\,{\AA} to 9.15~eV\,{\AA}~\cite{bordoloi2024}.

{\bf Strong SOC boosts the gate frequency of Ge hole spin qubit}

Once we obtain the linear Rashba parameters $\alpha_{\text{1}}$ and $\alpha_{\text{2}}$, it is straightforward to assess the Rabi frequency $f_{\text{R}}$ for planar hole spin qubits confined in QDs hosted by the 2D QW via an EDSR spin manipulation. Fig.~\ref{fig:fig2}c schematically depicts a planar hole state laterally confined by a gate-defined anisotropic harmonic potential $V(x',y')=\frac{\hbar^2}{2r_{\parallel}^4}(\frac{x'^2}{m_{x'}}+\frac{y'^2}{m_{y'}})$, where $r_{\parallel}$ is the effective lateral radius of the QD and $m_{x'}$ and $m_{y'}$ the in-plane effective mass of HH1 along the [110] and [$\overline{1}1$0] directions, respectively. To simplify the following arguments, we consider the alternating electric field $E_{\text{ac}}\cos\omega_{\text{ac}} t$ and magnetic field applied both along the $y'=[\overline{1}10]$ direction to drive the EDSR mediated by SOC, where the resonance of Rabi oscillation occurs if the frequency $\omega_{\text{ac}}$ matches the Larmor frequency $\omega_{\text{L}}$ of the qubit~\cite{Golovach_PRB2006}. Regarding most experiments are conducted at a constant Larmor frequency $\omega_\text{L}$ instead of a constant magnetic field, the Rabi frequency \yxadd{under the rotating-wave approximation (RWA)} is obtained for a constant $\omega_{\text{L}}$ as follows (see Supplementary Note 1)
\begin{equation}\label{Rabi}
f_{\text{R}}=\frac{\omega_{\text{ac}}}{2\pi\hbar^2}m_{y'}|\alpha_\text{1} -\alpha_\text{2}|\delta d_{\text{ac}},
\end{equation}
where $\delta d_{\text{ac}}=eE_{\text{ac}}m_{y'}r_{\parallel}^4/\hbar^2$ is the displacement of QD driven by the alternating electric field $E_{\text{ac}}$. Fig.~\ref{fig:biax}a shows the predicted Rabi frequency $f_{\text{R}}$ as a function of uniaxial strain $\epsilon_{\text{u}}$ according to equation~(\ref{Rabi}) for a hole spin qubit confined in a 120-ML thick Ge QW with $E_z=100$~kV\,cm$^{-1}$,  $r_{\parallel}= 50~\text{nm}$, $f_{\text{L}}=\omega_\text{L}/2\pi=5$~GHz, and $E_{\text{ac}}=0.02$~mV\,nm$^{-1}$ following the experimental setup~\cite{Abadillo-Uriel2023}. It shows that both compressive and tensile uniaxial strains increase $f_{\text{R}}$ linearly at a large rate. Specifically, $f_{\text{R}}$ reaches an ultrafast speed of 8~GHz at $\epsilon_{\text{u}}=-0.8\%$ and 6.5~GHz at $\epsilon_{\text{u}}=0.8\%$, exceeding the record 1.2~GHz speed of Ge hole spin qubit confined in one-dimensional Ge/Si hut wires~\cite{Liu2023}. Interestingly, compressive uniaxial strain raises $f_{\text{R}}$ rapidly at an even larger rate than the tensile uniaxial strain, even though the former enhances $|\alpha_{\text{1}}-\alpha_{\text{2}}|$ less efficiently than the latter (see Fig.~\ref{fig:fig2}g). This abnormal behavior results from the monotonic reduction of $m_{y'}$ from $0.203$ $m_0$ to $0.080$ $m_0$ as changing the uniaxial strain $\epsilon_{\text{u}}$ from $-0.8\%$ (compressive) through $0$ (no uniaxial strain) to $+0.8\%$ (tensile), as shown in Fig.~\ref{fig:biax}b.  Because Rabi frequency $f_{\text{R}}$ is linearly proportional to $m_{y'}$ (equation~(\ref{Rabi})), heavier $m_{y'}$ compensates the smaller $|\alpha_{\text{1}}-\alpha_{\text{2}}|$ for compressive strain relative to tensile strain.

Fig.~\ref{fig:biax}c shows that the corresponding Rabi frequency of hole spin qubits rises rapidly by reducing the biaxial strain and exceeds 20~GHz when the magnitude of the residual compressive biaxial strain reduces to less than 0.2\%. We can further raise $f_{\text{R}}$ to 30~GHz by increasing the tensile uniaxial strain in the case of an extremely small biaxial strain of $\epsilon_{\text{b}}=-0.1\%$, as shown in Fig.~\ref{fig:biax}d. A small compressive uniaxial strain can even boost $f_{\text{R}}$ to 40~GHz. Note that the proposed Rabi frequency of 40~GHz significantly surpasses the maximum ($\sim$10~GHz~\cite{Hendrickx2020, Liu2023}) that current arbitrary waveform generators in state-of-the-art experiments can support, shifting the breakthrough point for further acceleration from enhancing linear SOC to improving classical control.

{\bf Strong SOC allows low critical field for ``sweet spots''}

\begin{figure}
  \centering
  \includegraphics[width=\linewidth]{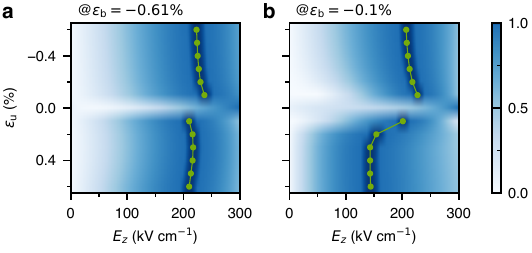}
  \caption{\textbf{Small biaxial strain enables low electric field for ``sweet spots''.} Linear SOC parameters $|\alpha_{1}-\alpha_{2}|$ as a function of electric field $E_z$ and uniaxial strain $\epsilon_{\text{u}}$ along the [110] direction under two different fixed biaxial strain: (\textbf{a}) $\epsilon_{\text{b}}=-0.61\%$, and (\textbf{b}) $\epsilon_{\text{b}}=-0.1\%$. The linear SOC parameters $|\alpha_1 - \alpha_2|$ are normalized to their maximum value for each uniaxial strain to highlight relative changes and ``sweet spots''. Green points indicate the estimated locations of the ``sweet spots'', where the linear SOC parameters $|\alpha_1-\alpha_2|$ reach maximum against the electric field $E_z$. Darker regions indicate the approximate ranges where the relative change of linear SOC parameters is less than 1\%.}
  \label{fig:sweet}
\end{figure}

Although considerable acceleration of qubit control is achieved, the substantially enhanced linear SOC is also expected to amplify the decoherence caused by a stronger coupling to charge noise. Fortunately, this type of decoherence can be significantly suppressed by operating at ``sweet spots'' where the derivative of Rabi frequency $f_{\rm R}$ with respect to $E_z$ vanishes~\cite{Wang2021}. Fig.~\ref{fig:fig2}f shows that both linear Rashba parameters $\alpha_{\text{1}}$ and $\alpha_{\text{2}}$ exhibit a maximum with increasing $E_z$, providing a sweet spot to suppress the decoherence caused by a related type of charge noise. Specifically,  for $\epsilon_{\text{u}}=0.4\%$, $|\alpha_{\text{1}}-\alpha_{\text{2}}|$ achieves the maximum of 204~meV\,{\AA} at $E_z=225$~kV\,cm$^{-1}$, giving rise to a zero derivative for the Rabi frequency $f_{\text{R}}$ with a critical electric field of $E_z=225$~kV\,cm$^{-1}$.  

However, it might be unfeasible to reach electric fields larger than 200~kV\,cm$^{-1}$ in Si-based devices. To explore the lower critical electric fields, we examine the ``sweet spots'' under different strains. Fig.~\ref{fig:sweet}a shows that under a biaxial strain $\epsilon_{\text{b}}=-0.61\%$, increasing the magnitude of the uniaxial strain alone fails to lower the critical electric fields effectively. However, if we reduce the biaxial strain to  $\epsilon_{\text{b}}=-0.1\%$, Fig.~\ref{fig:sweet}b shows that a small tensile strain could lower the critical electric field to around 150~kV\,cm$^{-1}$, which is more accessible in experiments, resulting from an extremely strong direct linear Rashba SOC (see Fig.~\ref{fig:fig2}i). We also note that the derivative of the linear SOC parameters with respect to uniaxial strain tends to zero at $\epsilon_{\text{u}}=0.2\%$, which relaxes the stringent requirements for precise uniaxial strain control. At this set combination of biaxial strain and uniaxial strain, the ``sweet spot'' is estimated to be located at 152~kV\,cm$^{-1}$, with the linear SOC parameter $|\alpha_1-\alpha_2|$ estimated to be 598~meV\,{\AA} and the Rabi frequency reaches 33.7~GHz, one order of magnitude faster than all types of known qubits~\cite{Anferov2024, bluvstein2022, Vallabhapurapu2023, Ha2022, Liu2023}. 

{\bf Strong SOC offers new qubit control regimes}

\yxadd{The RWA used in equation~(\ref{Rabi}) to estimate $f_{\text{R}}$ will become invalid once $f_{\text{R}}$ approaches $f_{\text{L}}$, which is believed to reduce the fidelity of single qubit gates or even hinder the gates themselves.
However, the high fidelity of single qubit gates can still be achieved by optimizing the pulse shape, as demonstrated in NV centers by a chopped random basis quantum optimization algorithm~\cite{scheuer2014} or by optimized resonant offset-sine drive pulses~\cite{yudilevich2023}, and in trapped ion qubits by mode-locked laser pulses~\cite{campbell2010}. In principle, these optimization strategies are applicable to any qubit platforms, including planar Ge hole spin qubits, as has been demonstrated in theory by using unitary perturbation theory analysis and numerical optimization of pulse parameters~\cite{ahn2024}. Additionally, experiments have shown that the departure from simple sinusoidal oscillations does not prevent gate operations~\cite{fuchs2009}.}

\begin{figure}
  \centering
  \includegraphics[width=\linewidth]{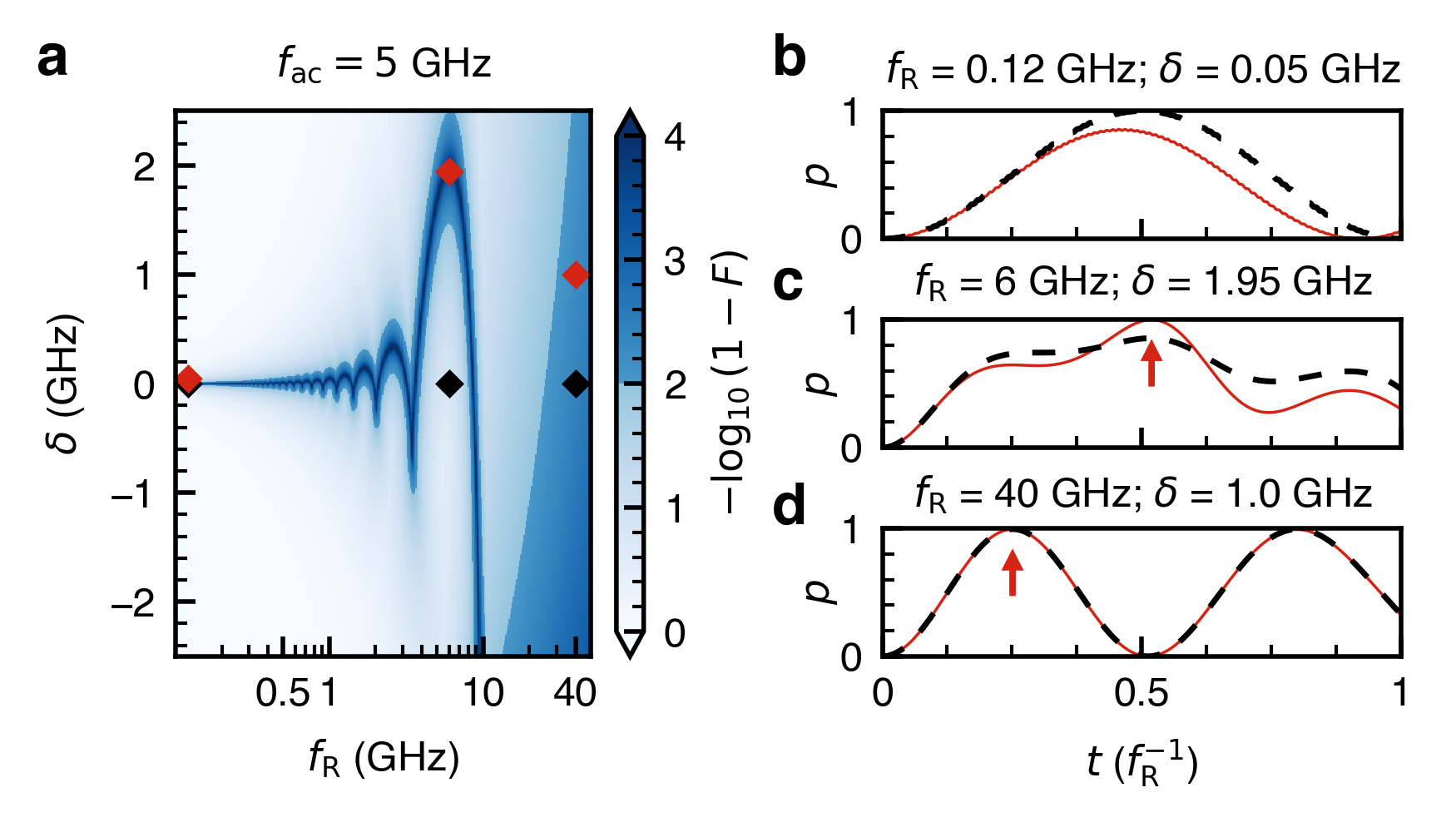}
  \caption{\textbf{Fidelity and dynamics of spin flipping of hole spin qubit.} (\textbf{a}) Predicted fidelity $F$ of spin flipping as a function of Rabi frequency $f_{\text{R}}$ and detuning $\delta=f_{\text{L}}-f_{\text{ac}}$ for $f_{\text{ac}}$ being set to 5~GHz. Darker regions represent $F>99\%$ ($-\log{(1-F)} > 2$). (\textbf{b}-\textbf{d}) The possibility of spin flipping $p$ as a function of time $t$ for (\textbf{b}) $f_{\text{R}}=0.12~\text{GHz}$ and detuning $\delta=-0.05~\text{GHz}$; (\textbf{c}) $f_{\text{R}}=6~\text{GHz}$, $\delta=1.95~\text{GHz}$; and (\textbf{d}) $f_{\text{R}}=40~\text{GHz}$, $\delta= 1.0~\text{GHz}$ (three sets of $f_{\text{R}}$ and $\delta$ are marked in (\textbf{a})). We compare them with the $\delta=0$ resonant cases. The first high-fidelity spin flipping in the detuned cases is marked by red arrows.}
  \label{fig:strdri}
\end{figure}

 \yxadd{To further demonstrate the advantages of strong driving, we explore the effect of detuning on the gate fidelity $F$ of planer Ge hole qubits by studying the motion of the spin subject to a strong SOC based on the effective Hamiltonian acting on spin rotation
  \begin{equation}\label{eq:heff}
    H_{\text{eff}} = \frac{1}{2} h f_{\text{L}} \sigma_{y'} + h f_{\text{R}} \cos(2\pi f_{\text{ac}} t) \sigma_{x'},
  \end{equation}
where $x'$ and $y'$ corresponds to $[110]$ and $[\bar{1}10]$ direction. For the sake of consistency throughout the paper, here, we continue to use the nominal $f_{\text{R}}$ in equation~(\ref{Rabi}) to characterize the SOC-driven effective magmetic field strength. Starting from one eigenstate of $\sigma_{y'}$, i.e., $\ket{\psi(t=0)}=\ket{0}$, the probability of spin flipping is defined as $p(t)=|\braket{\psi(t)}{1}|^2$ with $\ket{1}$ the other eigenstate. Assuming the electric field can be turned on and off abruptly and ignoring the trailing dynamics after turning off the pulse~\cite{fuchs2009}, the fidelity $F$ is defined as the maximum probability of spin flipping in a gate period of $T_0=f_{\text{R}}^{-1}$: $F=\max_{0\le t \le T_0} p(t)$. Fig.~\ref{fig:strdri}a shows the predicted $F$ as a function of $f_{\text{R}}$ and detuning $\delta=f_{\text{L}}-f_{\text{ac}}$ for a constant $f_{\text{ac}}$ of 5 GHz. One can observe that the spin precessing is well described by the RWA for $f_{\text{R}} \ll f_{\text{L}}$, where a moderate off-resonant driving will considerably reduce the fidelity (see Fig.~\ref{fig:strdri}b). As $f_{\text{R}}$ increases, the high-fidelity curve begins to oscillate with increasing amplitude, which can be approximately described by $\delta/f_{\text{ac}}=0.5(f_{\text{R}}/f_{\text{ac}})^2[\vert \cos(2\pi{f_{\text{ac}}}T_{\text{flip}})\vert-\phi]$ with phase $\phi\approx 1/\pi$ and spin-flipping time $T_{\text{flip}}\approx (2f_{\text{R}})^{-1}$. Consequently, the detuning $\delta$ must be finely adjusted to maintain a high fidelity (see Fig.~\ref{fig:strdri}c). Interestingly, even though high fidelity spin flipping is unaccessible for $10 < f_{\text{R}} \lesssim 25$ GHz without other optimization strategies, the achievement of high fidelity spin flipping become insensitive to detuning fluctuations when $f_{\text{R}} \gg f_{\text{L}}$. This occurs because the SOC-induced effective magnetic field $hf_{\text{R}}$ dominates over the external field $\frac{1}{2} hf_{\text{L}}$, and $2\pi f_{\text{ac}}T_0\ll 1$, leading to $H_{\text{eff}} \approx h f_{\text{R}} \cos(2\pi f_{\text{ac}} t) \sigma_{x'} \approx hf_{\text{R}}\sigma_{x'}$ (equation (\ref{eq:heff})) and an actual Rabi frequency of $2f_{\text{R}}$ (see Fig.~\ref{fig:strdri}d). 
In addition to being immune to frequency fluctuations, this insensitivity to detuning enables the simultaneous control of multiple qubits with a unified signal, which is favorable to potentially reducing calibration and circuit design overhead. Furthermore, since the required electric field pulse duration is shorter than the ac period, the control signal effectively operates at baseband, thus reducing heating effects. This regime can also be found in the scenarios with a longer gate period $T_0 = 2f_{\text{R}}^{-1}$ or with a fixed $f_{\text{L}}=5~\text{GHz}$ instead of fixed $f_{\text{ac}}$ (see Supplementary Note 5). These results demonstrate that extremely strong SOC offers a robust, high-speed control regime immune to conventional noise sources.}

{\bf Discussion}

\yxadd{The strong SOC usually reduces the energy separation to excited states, which causes more leakage out of the qubit encoding subspace, thus lowering the fidelity of all operations. Numerical simulations confirm that the energy gap to the first orbital excited state drops from 1.5~meV to above 0.8~meV (9.2~K), which is still high enough compared to standard sub-Kelvin operation temperatures (see Supplementary Note 6).}

While we have exclusively focused on the uniaxial strain along the [110] direction, ones applied along other directions, such as $\epsilon_{[100]}$, are also applicable to significantly enhance the linear Rashba SOC by lowering the QW symmetry to allow strong HH-LH mixing because of a finite $M_{\epsilon}$ presented in the BP Hamiltonian ($\epsilon_{xx}\neq\epsilon_{yy}$ and $\epsilon_{xy}=0$). Fig.~S3 shows the results for the case of $\epsilon_{[100]}$.

In addition to the superfast single qubit operation, the enhanced linear SOC is so strong that it makes Ge 2DHGs suitable for many spintronic applications. For example, it enables ultra-strong and even deep-strong spin-photon coupling~\cite{forn-diaz2019}, which is crucial to circuit quantum electrodynamics and photon-mediated long-range coupling between spin qubits~\cite{niemczyk2010}, yet so far unattained in planar Ge hole spin qubits. It also reduces the spin precession length from several $\mu$m to tens of nm for nanoscale spin field-effect transistor (SpinFET)~\cite{Datta1990}.

In conclusion, we have demonstrated that uniaxial strain provides a particularly effective tool for accelerating the all-electrical manipulation of planar Ge hole spin qubits via enhancing the linear Rashba SOC of 2D hole states.  We demonstrate that a small uniaxial strain, available within existing semiconductor technology, can boost  $f_\text{R}$ over 40~GHz in a vanishing biaxial strain of -0.1\%. The peak of the linear SOC parameters with respect to uniaxial strain $\epsilon_{\text{u}}$, the emergence of ``sweet spot'' due to the maximum of SOC strength against the vertical electric field, and the new qubit control regime that is insensitive to detuning together relax the stringent requirement during the device production and spin manipulation process. Therefore, these findings offer a simple way to develop superfast coherent control of planar Ge hole spin qubits towards large-scale quantum computing and planar Ge hole-based spintronics.

\bibliography{Strain_enhanced_Rashba} 

@Article{Zutic_RMP2004,
  Title                    = {Spintronics: Fundamentals and applications},
  Author                   = {Igor \v{Z}uti\'{c} and J. Fabian and S. D. Sarma},
  Journal                  = {Rev. Mod. Phys.},
  Year                     = {2004},
  Pages                    = {323},
  Volume                   = {76}
}

@Article{Abadillo-Uriel2023,
  Title                    = {Hole-Spin Driving by Strain-Induced Spin-Orbit Interactions},
  Author                   = {Abadillo-Uriel, Jos\'e Carlos and Rodr\'{\i}guez-Mena, Esteban A. and Martinez, Biel and Niquet, Yann-Michel},
  Journal                  = {Phys. Rev. Lett.},
  Year                     = {2023},

  Month                    = {Sep},
  Pages                    = {097002},
  Volume                   = {131},

  Doi                      = {10.1103/PhysRevLett.131.097002},
  Issue                    = {9},
  Numpages                 = {7},
  Publisher                = {American Physical Society},
}

@Article{Ast_PRL2007,
  Title                    = {Giant spin splitting through surface alloying},
  Author                   = {C. R. Ast and J. Henk and A. Ernst and L. Moreschini and M. C. Falub and D. Pacile and P. Bruno and K. Kern and M. Grioni},
  Journal                  = {Phys. Rev. Lett.},
  Year                     = {2007},
  Pages                    = {186807},
  Volume                   = {98}
}

@Article{Awschalom2013,
  Title                    = {Quantum Spintronics: Engineering and Manipulating Atom-Like Spins in Semiconductors},
  Author                   = {Awschalom, David D. and Bassett, Lee C. and Dzurak, Andrew S. and Hu, Evelyn L. and Petta, Jason R.},
  Journal                  = {Science},
  Year                     = {2013},
  Number                   = {6124},
  Pages                    = {1174--1179},
  Volume                   = {339},
  Publisher                = {American Association for the Advancement of Science}
}

@Article{Bernevig2005,
  Title                    = {Intrinsic Spin {{Hall}} Effect in the Two-Dimensional Hole Gas},
  Author                   = {B. Andrei Bernevig and Shou-Cheng Zhang},
  Journal                  = {Phys. Rev. Lett.},
  Year                     = {2005},
  Pages                    = {016801},
  Volume                   = {95}
}

@Article{Bernevig2006,
  Title                    = {Quantum spin {{Hall}} effect and topological phase transition in {{HgTe}} quantum wells},
  Author                   = {Bernevig, B. A. and Hunghes, T. L. and Zhang, S. -C.},
  Journal                  = {Science},
  Year                     = {2006},
  Pages                    = {1757-1761},
  Volume                   = {314}
}

@article{borsoi2024,
  title = {Shared Control of a 16 Semiconductor Quantum Dot Crossbar Array},
  author = {Borsoi, Francesco and Hendrickx, Nico W. and John, Valentin and Meyer, Marcel and Motz, Sayr and {van Riggelen}, Floor and Sammak, Amir and {de Snoo}, Sander L. and Scappucci, Giordano and Veldhorst, Menno},
  year = 2024,
  month = jan,
  journal = {Nat. Nanotechnol.},
  volume = {19},
  number = {1},
  pages = {21--27},
  publisher = {Nature Publishing Group},
  issn = {1748-3395},
  doi = {10.1038/s41565-023-01491-3},
  urldate = {2024-04-18},
  copyright = {2023 The Author(s)}
}

@Article{Bosco2021,
  Title                    = {Squeezed hole spin qubits in Ge quantum dots with ultrafast gates at low power},
  Author                   = {Bosco, Stefano and Benito, M\'onica and Adelsberger, Christoph and Loss, Daniel},
  Journal                  = {Phys. Rev. B},
  Year                     = {2021},

  Month                    = {Sep},
  Pages                    = {115425},
  Volume                   = {104},
  Issue                    = {11},
  Numpages                 = {6},
  Publisher                = {American Physical Society}
}

@Article{EDSR2007,
  Title                    = {Electric Dipole Spin Resonance for Heavy Holes in Quantum Dots},
  Author                   = {Denis V. Bulaev and Daniel Loss},
  Journal                  = {Phys. Rev. Lett.},
  Year                     = {2007},
  Pages                    = {097202},
  Volume                   = {98}
}

@Article{Corley-Wiciak2023,
  Title                    = {Nanoscale Mapping of the {{3D}} Strain Tensor in a Germanium Quantum Well Hosting a Functional Spin Qubit Device},
  Author                   = {Corley-Wiciak, Cedric and Richter, Carsten and Zoellner, Marvin H. and Zaitsev, Ignatii and Manganelli, Costanza L. and Zatterin, Edoardo and Sch眉lli, Tobias U. and Corley-Wiciak, Agnieszka A. and Katzer, Jens and Reichmann, Felix and Klesse, Wolfgang M. and Hendrickx, Nico W. and Sammak, Amir and Veldhorst, Menno and Scappucci, Giordano and Virgilio, Michele and Capellini, Giovanni},
  Journal                  = {ACS Appl. Mater. Interfaces},
  Year                     = {2023},

  Month                    = jan,
  Number                   = {2},
  Pages                    = {3119--3130},
  Volume                   = {15},

  __markedentry            = {[dlk:]},
  Booktitle                = {ACS Applied Materials \& Interfaces},
  Comment                  = {doi: 10.1021/acsami.2c17395},
  Doi                      = {10.1021/acsami.2c17395},
  File                     = {Corley-Wiciak2023.pdf:pdf\\Corley-Wiciak2023.pdf:PDF},
  ISSN                     = {1944-8244},
  Owner                    = {dlk},
  Publisher                = {American Chemical Society},
  Timestamp                = {2025.02.24},
}

@Article{Datta1990,
  Title                    = {Electronic analog of the eiectro-optic modulator},
  Author                   = {Supriyo Datta and Biswajit Das},
  Journal                  = {Appl. Phys. Lett.},
  Year                     = {1990},
  Pages                    = {665--667},
  Volume                   = {56}
}

@article{frantzeskakis2008,
  title = {Tunable {{Spin Gaps}} in a {{Quantum-Confined Geometry}}},
  author = {Frantzeskakis, Emmanouil and Pons, St{\'e}phane and Mirhosseini, Hossein and Henk, J{\"u}rgen and Ast, Christian R. and Grioni, Marco},
  year = 2008,
  month = nov,
  journal = {Physical Review Letters},
  volume = {101},
  number = {19},
  pages = {196805},
  issn = {0031-9007, 1079-7114},
  doi = {10.1103/PhysRevLett.101.196805},
  urldate = {2026-06-22},
  copyright = {http://link.aps.org/licenses/aps-default-license}
}

@article{he2008,
  title = {Spin {{Polarization}} of {{Quantum Well States}} in {{Ag Films Induced}} by the {{Rashba Effect}} at the {{Surface}}},
  author = {He, Ke and Hirahara, Toru and Okuda, Taichi and Hasegawa, Shuji and Kakizaki, Akito and Matsuda, Iwao},
  year = 2008,
  month = sep,
  journal = {Physical Review Letters},
  volume = {101},
  number = {10},
  pages = {107604},
  issn = {0031-9007, 1079-7114},
  doi = {10.1103/PhysRevLett.101.107604},
  urldate = {2026-06-22},
  copyright = {http://link.aps.org/licenses/aps-default-license}
}

@Article{Golovach_PRB2006,
  Title                    = {Electric-dipole-induced spin resonance in quantum dots},
  Author                   = {Vitaly N. Golovach and Massoud Borhani and Daniel Loss},
  Journal                  = {Phys. Rev. B},
  Year                     = {2006},
  Pages                    = {165319},
  Volume                   = {74}
}

@Article{Ha2022,
  Title                    = {A Flexible Design Platform for {{Si/SiGe}} Exchange-Only Qubits with Low Disorder},
  Author                   = {Ha, Wonill and Ha, Sieu D. and Choi, Maxwell D. and Tang, Yan and Schmitz, Adele E. and Levendorf, Mark P. and Lee, Kangmu and Chappell, James M. and Adams, Tower S. and Hulbert, Daniel R. and Acuna, Edwin and Noah, Ramsey S. and Matten, Justine W. and Jura, Michael P. and Wright, Jeffrey A. and Rakher, Matthew T. and Borselli, Matthew G.},
  Journal                  = {Nano Letters},
  Year                     = {2022},
  Number                   = {3},
  Pages                    = {1443-1448},
  Volume                   = {22}
}

@Article{Hendrickx2020,
  Title                    = {Fast two-qubit logic with holes in germanium},
  Author                   = {N. W. Hendrickx and D. P. Franke and A. Sammak and G. Scappucci and M. Veldhorst},
  Journal                  = {Nature},
  Year                     = {2020},
  Pages                    = {487},
  Volume                   = {577}
}

@Article{Hendrickx2021,
  Title                    = {A four-qubit germanium quantum processor},
  Author                   = {Hendrickx, Nico W. and Lawrie, William I. L. and Russ, Maximilian and van Riggelen, Floor and de Snoo, Sander L. and Schouten, Raymond N. and Sammak, Amir and Scappucci, Giordano and Veldhorst, Menno},
  Journal                  = {Nature},
  Year                     = {2021},

  Month                    = mar,
  Number                   = {7851},
  Pages                    = {580--585},
  Volume                   = {591},
  Refid                    = {Hendrickx2021}
}

@Article{King_PRL2011,
  Title                    = {Large tunable {{Rashba}} spin splitting of a two-dimensional electron gas in $\mathrm{Bi_2Se_3}$},
  Author                   = {P. D. C. King and R. C. Hatch and M. Bianchi and R. Ovsyannikov and C. Lupulescu and G. Landolt and B. Slomski and J. H. Dil and D. Guan and J. L. Mi and E. D. L. Rienks and J. Fink and A. Lindblad and S. Svensson and S. Bao and G. Balakrishnan and B. B. Iversen and J. Osterwalder and W. Eberhardt and F. Baumberger and P. Hofmann},
  Journal                  = {Phys. Rev. Lett.},
  Year                     = {2011},
  Pages                    = {096802},
  Volume                   = {107}
}

@Article{Kloeffel2011,
  Title                    = {Strong spin-orbit interaction and helical hole states in {{Ge/Si}} nanowires},
  Author                   = {Kloeffel, Christoph and Trif, Mircea and Loss, Daniel},
  Journal                  = {Phys. Rev. B},
  Year                     = {2011},

  Month                    = {Nov},
  Pages                    = {195314},
  Volume                   = {84},

  Issue                    = {19},
  Numpages                 = {8},
  Publisher                = {American Physical Society}
}

@Article{Koroteev_PRL2004,
  Title                    = {Strong spin-orbit splitting on {{Bi}} surfaces},
  Author                   = {Y. M. Koroteev and G. Bihlmayer and J. E. Gayone and E. V. Chulkov and S. Blugel and P. M. Echenique and P. Hofmann},
  Journal                  = {Phys. Rev. Lett.},
  Year                     = {2004},
  Pages                    = {046403},
  Volume                   = {93}
}

@Article{Laroche2016,
  Title                    = {Magneto-transport analysis of an ultra-low-density two-dimensional hole gas in an undoped strained {{Ge/SiGe}} heterostructure},
  Author                   = {Laroche, D. and Huang, S.-H. and Chuang, Y. and Li, J.-Y. and Liu, C. W. and Lu, T. M.},
  Journal                  = {Applied Physics Letters},
  Year                     = {2016},

  Month                    = {06},
  Number                   = {23},
  Pages                    = {233504},
  Volume                   = {108},
  Doi                      = {10.1063/1.4953399},
  ISSN                     = {0003-6951},
}

@Article{LaShell_PRL1996,
  Title                    = {Spin splitting of an {{Au(111)}} surface state band observed with angle resolved photoelectron spectroscopy},
  Author                   = {S. LaShell and B. A. McDougall and E. Jensen},
  Journal                  = {Phys. Rev. Lett.},
  Year                     = {1996},
  Pages                    = {3419},
  Volume                   = {77}
}

@Article{Lee2004,
  Title                    = {{Strained Si, SiGe, and Ge channels for high-mobility metal-oxide-semiconductor field-effect transistors}},
  Author                   = {Lee, Minjoo L. and Fitzgerald, Eugene A. and Bulsara, Mayank T. and Currie, Matthew T. and Lochtefeld, Anthony},
  Journal                  = {Journal of Applied Physics},
  Year                     = {2004},

  Month                    = {12},
  Number                   = {1},
  Pages                    = {011101},
  Volume                   = {97},
  Doi                      = {10.1063/1.1819976},
  ISSN                     = {0021-8979},
}

@Article{Liu2023,
  Title                    = {Ultrafast and Electrically Tunable Rabi Frequency in a Germanium Hut Wire Hole Spin Qubit},
  Author                   = {Liu, He and Wang, Ke and Gao, Fei and Leng, Jin and Liu, Yang and Zhou, Yu-Chen and Cao, Gang and Wang, Ting and Zhang, Jianjun and Huang, Peihao and Li, Hai-Ou and Guo, Guo-Ping},
  Journal                  = {Nano Lett.},
  Year                     = {2023},

  Month                    = may,
  Number                   = {9},
  Pages                    = {3810--3817},
  Volume                   = {23},

  Booktitle                = {Nano Letters},
  Comment                  = {doi: 10.1021/acs.nanolett.3c00213},
  Doi                      = {10.1021/acs.nanolett.3c00213},
  ISSN                     = {1530-6984},
  Publisher                = {American Chemical Society},
}

@Article{LiuyangPRB2022,
  Title                    = {Emergent linear {{Rashba}} spin-orbit coupling offers fast manipulation of hole-spin qubits in germanium},
  Author                   = {Liu, Yang and Xiong, Jia-Xin and Wang, Zhi and Ma, Wen-Long and Guan, Shan and Luo, Jun-Wei and Li, Shu-Shen},
  Journal                  = {Phys. Rev. B},
  Year                     = 2022,
  Pages                    = 075313,
  Volume                   = 105
}

@Article{Lodari2022,
  Title                    = {Lightly strained germanium quantum wells with hole mobility exceeding one million},
  Author                   = {Lodari, M. and Kong, O. and Rendell, M. and Tosato, A. and Sammak, A. and Veldhorst, M. and Hamilton, A. R. and Scappucci, G.},
  Journal                  = {Applied Physics Letters},
  Year                     = {2022},

  Month                    = {03},
  Number                   = {12},
  Pages                    = {122104},
  Volume                   = {120},

  __markedentry            = {[dlk:]},
  Doi                      = {10.1063/5.0083161},
  ISSN                     = {0003-6951},
  Owner                    = {dlk},
  Timestamp                = {2025.02.24},
}

@Article{Luo_PRB2015,
  Title                    = {Supercoupling between heavy-hole and light-hole states in nanostructures},
  Author                   = {Luo, Jun-Wei and Bester, Gabriel and Zunger, Alex},
  Journal                  = {Phys. Rev. B},
  Year                     = {2015},

  Month                    = {Oct},
  Pages                    = {165301},
  Volume                   = {92},

  Issue                    = {16},
  Numpages                 = {9},
  Publisher                = {American Physical Society}
}

@Article{Luo_PRL2009,
  Title                    = {Full-zone spin splitting for electrons and holes in bulk {{GaAs}} and {{GaSb}}},
  Author                   = {Jun-Wei Luo and Gabriel Bester and Alex Zunger},
  Journal                  = {Phys. Rev. Lett.},
  Year                     = {2009},
  Pages                    = {056405},
  Volume                   = {102}
}

@Article{Luo2010,
  Title                    = {Discovery of a Novel Linear-in-$k$ Spin Splitting for Holes in the {{2D}} $\mathrm{GaAs}/\mathrm{AlAs}$ System},
  Author                   = {Luo, Jun-Wei and Chantis, Athanasios N. and van Schilfgaarde, Mark and Bester, Gabriel and Zunger, Alex},
  Journal                  = {Phys. Rev. Lett.},
  Year                     = {2010},

  Month                    = {Feb},
  Pages                    = {066405},
  Volume                   = {104},

  Issue                    = {6},
  Numpages                 = {4},
  Publisher                = {American Physical Society}
}

@Article{Luo2017,
  Title                    = {Rapid Transition of the Hole Rashba Effect from Strong Field Dependence to Saturation in Semiconductor Nanowires},
  Author                   = {Luo, Jun-Wei and Li, Shu-Shen and Zunger, Alex},
  Journal                  = {Phys. Rev. Lett.},
  Year                     = {2017},

  Month                    = {Sep},
  Pages                    = {126401},
  Volume                   = {119},

  Issue                    = {12},
  Numpages                 = {6},
  Publisher                = {American Physical Society}
}

@Article{Luo2011,
  Title                    = {Absence of intrinsic spin splitting in one-dimensional quantum wires of tetrahedral semiconductors},
  Author                   = {Luo, Jun-Wei and Zhang, Lijun and Zunger, Alex},
  Journal                  = {Phys. Rev. B (R)},
  Year                     = {2011},

  Month                    = {Sep},
  Pages                    = {121303},
  Volume                   = {84},

  Issue                    = {12},
  Numpages                 = {4},
  Publisher                = {American Physical Society}
}

@Article{LuoJW_PRL2010,
  Title                    = {Design principles and coupling mechanisms in the 2D quantum well topological insulator {{HgTe/CdTe}}},
  Author                   = {Jun-Wei Luo and Alex Zunger},
  Journal                  = {Phys. Rev. Lett.},
  Year                     = {2010},
  Pages                    = {176805},
  Volume                   = {105}
}

@Article{Sarma2010,
  Title                    = {Majorana Fermions and a Topological Phase Transition in Semiconductor-Superconductor Heterostructures},
  Author                   = {R. M. Lutchyn and J. D. Sau and S. Das Sarma},
  Journal                  = {Phys. Rev. Lett.},
  Year                     = {2010},
  Pages                    = {077001},
  Volume                   = {105}
}

@Article{Manchon2015,
  Title                    = {New perspectives for {{Rashba}} spin-orbit coupling},
  Author                   = {A. Manchon and H. C. Koo and J. Nitta and S. M. Frolov and R. A. Duine},
  Journal                  = {Nat. Mater.},
  Year                     = {2015},
  Pages                    = {871-882},
  Volume                   = {14}
}

@Article{Maurand2016,
  Title                    = {A {{CMOS}} silicon spin qubit},
  Author                   = {Maurand, R. and Jehl, X. and Kotekar-Patil, D. and Corna, A. and Bohuslavskyi, H. and Lavieville, R. and Hutin, L. and Barraud, S. and Vinet, M. and Sanquer, M. and De Franceschi, S.},
  Journal                  = {Nat. Commun.},
  Year                     = {2016},
  Pages                    = {13575},
  Volume                   = {7}
}

@Article{Mourik2012,
  Title                    = {Signatures of {{Majorana}} fermions in hybrid superconductor-semiconductor nanowire devices},
  Author                   = {V. Mourik and K. Zuo and S. M. Frolov and S. R. Plissard and E. P. A. M. Bakkers and L. P. Kouwenhoven},
  Journal                  = {Science},
  Year                     = {2012},
  Pages                    = {1003},
  Volume                   = {336}
}

@Article{Zunger_JAP1998,
  Title                    = {Comparison of two methods for describing the strain profiles in quantum dots},
  Author                   = {C. Pryor and J. Kim and L. W. Wang and A. J. Williamson and A. Zunger},
  Journal                  = {J. Appl. Phys.},
  Year                     = {1998},
  Pages                    = {2548},
  Volume                   = {83}
}

@Article{Rodriguez-Mena2023,
  Title                    = {Linear-in-momentum spin orbit interactions in planar {{Ge/GeSi}} heterostructures and spin qubits},
  Author                   = {Rodr\'{\i}guez-Mena, Esteban A. and Abadillo-Uriel, Jos\'e Carlos and Veste, Ga\"etan and Martinez, Biel and Li, Jing and Skl\'enard, Beno\^{\i}t and Niquet, Yann-Michel},
  Journal                  = {Phys. Rev. B},
  Year                     = {2023},

  Month                    = {Nov},
  Pages                    = {205416},
  Volume                   = {108},

  Doi                      = {10.1103/PhysRevB.108.205416},
  Issue                    = {20},
  Numpages                 = {17},
  Publisher                = {American Physical Society},
}

@Article{Schliemann2003,
  Title                    = {Nonballistic Spin-Field-Effect Transistor},
  Author                   = {Schliemann, John and Egues, J. Carlos and Loss, Daniel},
  Journal                  = {Phys. Rev. Lett.},
  Year                     = {2003},

  Month                    = {Apr},
  Pages                    = {146801},
  Volume                   = {90},

  Issue                    = {14},
  Numpages                 = {4},
  Publisher                = {American Physical Society}
}

@Article{Sinova2004,
  Title                    = {Universal Intrinsic Spin {{Hall}} Effect},
  Author                   = {Sinova, Jairo and Culcer, Dimitrie and Niu, Q. and Sinitsyn, N. A. and Jungwirth, T. and MacDonald, A. H.},
  Journal                  = {Phys. Rev. Lett.},
  Year                     = {2004},

  Month                    = {Mar},
  Pages                    = {126603},
  Volume                   = {92},

  Issue                    = {12},
  Numpages                 = {4},
  Publisher                = {American Physical Society}
}

@Article{Terrazos2021,
  Title                    = {Theory of hole-spin qubits in strained germanium quantum dots},
  Author                   = {Terrazos, L. A. and Marcellina, E. and Wang, Zhanning and Coppersmith, S. N. and Friesen, Mark and Hamilton, A. R. and Hu, Xuedong and Koiller, Belita and Saraiva, A. L. and Culcer, Dimitrie and Capaz, Rodrigo B.},
  Journal                  = {Phys. Rev. B},
  Year                     = {2021},

  Month                    = {Mar},
  Pages                    = {125201},
  Volume                   = {103},

  Issue                    = {12},
  Numpages                 = {10},
  Publisher                = {American Physical Society}
}

@Article{Venitucci2018,
  Title                    = {Electrical manipulation of semiconductor spin qubits within the $g$-matrix formalism},
  Author                   = {Venitucci, Benjamin and Bourdet, L\'eo and Pouzada, Daniel and Niquet, Yann-Michel},
  Journal                  = {Phys. Rev. B},
  Year                     = {2018},

  Month                    = {Oct},
  Pages                    = {155319},
  Volume                   = {98},

  Doi                      = {10.1103/PhysRevB.98.155319},
  Issue                    = {15},
  Numpages                 = {17},
  Publisher                = {American Physical Society},
}

@Article{Wang2024,
  Title                    = {Operating semiconductor quantum processors with hopping spins},
  Author                   = {Chien-An Wang and Valentin John and Hanifa Tidjani and Cécile X. Yu and Alexander S. Ivlev and Corentin Déprez and Floor van Riggelen-Doelman and Benjamin D. Woods and Nico W. Hendrickx and William I. L. Lawrie and Lucas E. A. Stehouwer and Stefan D. Oosterhout and Amir Sammak and Mark Friesen and Giordano Scappucci and Sander L. de Snoo and Maximilian Rimbach-Russ and Francesco Borsoi and Menno Veldhorst},
  Journal                  = {Science},
  Year                     = {2024},
  Number                   = {6707},
  Pages                    = {447-452},
  Volume                   = {385},
  Doi                      = {10.1126/science.ado5915},
}

@Article{Wang_NC2022,
  Title                    = {Ultrafast coherent control of a hole spin qubit in a germanium quantum dot},
  Author                   = {Wang, Ke and Xu, Gang and Gao, Fei and Liu, He and Ma, Rong-Long and Zhang, Xin and Wang, Zhanning and Cao, Gang and Wang, Ting and Zhang, Jian-Jun and Culcer, Dimitrie and Hu, Xuedong and Jiang, Hong-Wen and Li, Hai-Ou and Guo, Guang-Can and Guo, Guo-Ping},
  Journal                  = {Nature Communications},
  Year                     = {2022},

  Month                    = jan,
  Number                   = {1},
  Pages                    = {206},
  Volume                   = {13},
  Refid                    = {Wang2022}
}

@Article{Wang1995,
  Title                    = {Local-density-derived semiempirical pseudopotentials},
  Author                   = {Wang, Lin-Wang and Zunger, Alex},
  Journal                  = {Phys. Rev. B},
  Year                     = {1995},

  Month                    = {Jun},
  Pages                    = {17398--17416},
  Volume                   = {51},

  Issue                    = {24},
  Numpages                 = {0},
  Publisher                = {American Physical Society}
}

@Article{Wang_JCP1994,
  Title                    = {Solving {{Schr{\"o}dinger's}} equation around a desired energy: Application to silicon quantum dots},
  Author                   = {Lin-Wang Wang and Alex Zunger},
  Journal                  = {J. Chem. Phys.},
  Year                     = 1994,
  Pages                    = 2394,
  Volume                   = 100
}

@Article{Wang2021,
  Title                    = {Optimal operation points for ultrafast, highly coherent {{Ge}} hole spin-orbit qubits},
  Author                   = {Wang, Zhanning and Marcellina, Elizabeth and Hamilton, Alex. R. and Cullen, James H. and Rogge, Sven and Salfi, Joe and Culcer, Dimitrie},
  Journal                  = {npj Quantum Information},
  Year                     = 2021,
  Number                   = 1,
  Pages                    = 54,
  Volume                   = 7,
  Refid                    = {Wang2021}
}

@Article{Watzinger2018,
  Title                    = {A germanium hole spin qubit},
  Author                   = {Watzinger, Hannes and Kukucka, Josip and Vukusic, Lada and Gao, Fei and Wang, Ting and Schaffler, Friedrich and Zhang, Jian-Jun and Katsaros, Georgios},
  Journal                  = {Nature Communications},
  Year                     = {2018},
  Number                   = {1},
  Pages                    = {3902},
  Volume                   = {9}
}

@Article{WangLW_PRB2000,
  Title                    = {Theoretical interpretation of the experimental electronic structure of lens-shaped self-assembled {{InAs/GaAs}} quantum dots},
  Author                   = {A. J. Williamson and L. W. Wang and A. Zunger},
  Journal                  = {Phys. Rev. B},
  Year                     = 2000,
  Pages                    = 12963,
  Volume                   = 62
}

@Article{DiXiao2010,
  Title                    = {Berry phase effects on electronic properties},
  Author                   = {Di Xiao and Ming-Che Chang and Qian Niu},
  Journal                  = {Rev. Mod. Phys.},
  Year                     = {2010},
  Pages                    = {1959-2007},
  Volume                   = {82}
}

@Article{Xiong2022,
  Title                    = {Orientation-dependent {{Rashba}} spin-orbit coupling of two-dimensional hole gases in semiconductor quantum wells: Linear or cubic},
  Author                   = {Xiong, Jia-Xin and Guan, Shan and Luo, Jun-Wei and Li, Shu-Shen},
  Journal                  = {Phys. Rev. B},
  Year                     = {2022},

  Month                    = {Mar},
  Pages                    = {115303},
  Volume                   = {105},

  Doi                      = {10.1103/PhysRevB.105.115303},
  Issue                    = {11},
  Numpages                 = {14},
  Publisher                = {American Physical Society},
}

@Article{Xiong_PRB2021,
  Title                    = {Emergence of the strong tunable linear {{Rashba}} spin-orbit coupling of two-dimensional hole gases in semiconductor quantum wells},
  Author                   = {Jia-Xin Xiong and Shan Guan and Jun-Wei Luo and Shu-Shen Li},
  Journal                  = {Phys. Rev. B},
  Year                     = {2021},
  Pages                    = {085309},
  Volume                   = {103}
}

@Book{Peter2005,
  Title                    = {Fundamentals of Semiconductors Physics and Materials Properties},
  Author                   = {P. Y. Yu and M. Cardona},
  Publisher                = {Springer},
  Year                     = {2005}
}

@Article{Luo_NC2013,
  Title                    = {Genetic design of enhanced valley splitting towards a spin qubit in silicon},
  Author                   = {L.-J. Zhang and J.-W. Luo and A. Saraiva and B. Koiller and A. Zunger},
  Journal                  = {Nat. Commun.},
  Year                     = {2013},
  Pages                    = {2396},
  Volume                   = {4}
}

@Article{Zunger2002,
  Title                    = {On the farsightedness (hyperopia) of the standard $k\cdot p$ method},
  Author                   = {Alex Zunger},
  Journal                  = {Phys. Stat. Sol. A},
  Year                     = {2002},
  Pages                    = {467--475},
  Volume                   = {190}
}

@article{Anferov2024,
  title = {Superconducting Qubits above 20 {{GHz}} Operating over 200 {{mK}}},
  author = {Anferov, Alexander and Harvey, Shannon P. and Wan, Fanghui and Simon, Jonathan and Schuster, David I.},
  journal = {PRX Quantum},
  volume = {5},
  issue = {3},
  pages = {030347},
  numpages = {19},
  year = {2024},
  month = {Sep},
  publisher = {American Physical Society},
  doi = {10.1103/PRXQuantum.5.030347},
}

@article{li2023,
  title = {Error per Single-Qubit Gate below $10^{-4}$ in a Superconducting Qubit},
  author = {Li, Zhiyuan and Liu, Pei and Zhao, Peng and Mi, Zhenyu and Xu, Huikai and Liang, Xuehui and Su, Tang and Sun, Weijie and Xue, Guangming and Zhang, Jing-Ning and Liu, Weiyang and Jin, Yirong and Yu, Haifeng},
  year = {2023},
  month = nov,
  journal = {npj Quantum Information},
  volume = {9},
  number = {1},
  pages = {111},
  issn = {2056-6387},
  doi = {10.1038/s41534-023-00781-x},
}

@article{bluvstein2022,
  title = {A Quantum Processor Based on Coherent Transport of Entangled Atom Arrays},
  author = {Bluvstein, Dolev and Levine, Harry and Semeghini, Giulia and Wang, Tout T. and Ebadi, Sepehr and Kalinowski, Marcin and Keesling, Alexander and Maskara, Nishad and Pichler, Hannes and Greiner, Markus and Vuletić, Vladan and Lukin, Mikhail D.},
  date = {2022-04},
  year = {2022},
  journal = {Nature},
  volume = {604},
  number = {7906},
  pages = {451--456},
  publisher = {Nature Publishing Group},
  issn = {1476-4687},
  doi = {10.1038/s41586-022-04592-6},
}

@article{Vallabhapurapu2023,
  title = {High-fidelity control of a nitrogen-vacancy-center spin qubit at room temperature using the sinusoidally modulated, always rotating, and tailored protocol},
  author = {Vallabhapurapu, Hyma H. and Hansen, Ingvild and Adambukulam, Chris and St\"ohr, Rainer and Denisenko, Andrej and Yang, Chih Hwan and Laucht, Arne},
  journal = {Phys. Rev. A},
  volume = {108},
  issue = {2},
  pages = {022606},
  numpages = {7},
  year = {2023},
  month = {Aug},
  publisher = {American Physical Society},
  doi = {10.1103/PhysRevA.108.022606},
}

@article{camenzind2022,
  title = {A Hole Spin Qubit in a Fin Field-Effect Transistor above 4 {{Kelvin}}},
  author = {Camenzind, Leon C. and Geyer, Simon and Fuhrer, Andreas and Warburton, Richard J. and Zumbühl, Dominik M. and Kuhlmann, Andreas V.},
  year = 2022,
  journal = {Nature Electronics},
  volume = 5,
  number = 3,
  pages = {178--183},
  publisher = {Nature Publishing Group},
  issn = {2520-1131},
  doi = {10.1038/s41928-022-00722-0},
}

@article{Nakajima2020,
  title = {Coherence of a Driven Electron Spin Qubit Actively Decoupled from Quasistatic Noise},
  author = {Nakajima, Takashi and Noiri, Akito and Kawasaki, Kento and Yoneda, Jun and Stano, Peter and Amaha, Shinichi and Otsuka, Tomohiro and Takeda, Kenta and Delbecq, Matthieu R. and Allison, Giles and Ludwig, Arne and Wieck, Andreas D. and Loss, Daniel and Tarucha, Seigo},
  journal = {Phys. Rev. X},
  volume = {10},
  issue = {1},
  pages = {011060},
  numpages = {11},
  year = {2020},
  month = {Mar},
  publisher = {American Physical Society},
  doi = {10.1103/PhysRevX.10.011060},
}

@article{lawrie2023,
  title = {Simultaneous Single-Qubit Driving of Semiconductor Spin Qubits at the Fault-Tolerant Threshold},
  author = {Lawrie, W. I. L. and {Rimbach-Russ}, M. and van Riggelen, F. and Hendrickx, N. W. and de Snoo, S. L. and Sammak, A. and Scappucci, G. and Helsen, J. and Veldhorst, M.},
  year = {2023},
  month = jun,
  journal = {Nature Communications},
  volume = {14},
  number = {1},
  pages = {3617},
  publisher = {Nature Publishing Group},
  issn = {2041-1723},
  doi = {10.1038/s41467-023-39334-3},
  urldate = {2025-04-25},
}

@article{campbell2010,
  title = {Ultrafast {{Gates}} for {{Single Atomic Qubits}}},
  author = {Campbell, W. C. and Mizrahi, J. and Quraishi, Q. and Senko, C. and Hayes, D. and Hucul, D. and Matsukevich, D. N. and Maunz, P. and Monroe, C.},
  year = {2010},
  month = aug,
  journal = {Phys. Rev. Lett.},
  volume = {105},
  number = {9},
  pages = {090502},
  publisher = {American Physical Society},
  doi = {10.1103/PhysRevLett.105.090502},
  urldate = {2025-05-16},
}

@article{howard2023,
  title = {Implementing Two-Qubit Gates at the Quantum Speed Limit},
  author = {Howard, Joel and Lidiak, Alexander and Jameson, Casey and Basyildiz, Bora and Clark, Kyle and Zhao, Tongyu and Bal, Mustafa and Long, Junling and Pappas, David P. and Singh, Meenakshi and Gong, Zhexuan},
  year = {2023},
  month = dec,
  journal = {Physical Review Research},
  volume = {5},
  number = {4},
  pages = {043194},
  publisher = {American Physical Society},
  doi = {10.1103/PhysRevResearch.5.043194},
  urldate = {2025-05-16},
}

@article{rower2024,
  title = {Suppressing {{Counter-Rotating Errors}} for {{Fast Single-Qubit Gates}} with {{Fluxonium}}},
  author = {Rower, David A. and Ding, Leon and Zhang, Helin and Hays, Max and An, Junyoung and Harrington, Patrick M. and Rosen, Ilan T. and Gertler, Jeffrey M. and Hazard, Thomas M. and Niedzielski, Bethany M. and Schwartz, Mollie E. and Gustavsson, Simon and Serniak, Kyle and Grover, Jeffrey A. and Oliver, William D.},
  year = {2024},
  month = dec,
  journal = {PRX Quantum},
  volume = {5},
  number = {4},
  pages = {040342},
  publisher = {American Physical Society},
  doi = {10.1103/PRXQuantum.5.040342},
  urldate = {2025-05-16},
}

@article{saner2023,
  title = {Breaking the {{Entangling Gate Speed Limit}} for {{Trapped-Ion Qubits}}},
  author = {Saner, S. and Bazavan, O. and Minder, M. and Drmota, P. and Webb, D. J. and Araneda, G. and Srinivas, R. and Lucas, D. M. and Ballance, C. J.},
  year = 2023,
  month = dec,
  journal = {Phys. Rev. Lett.},
  volume = 131,
  number = 22,
  pages = 220601,
  publisher = {Amer Physical Soc},
  address = {College Pk},
  issn = {0031-9007, 1079-7114},
  doi = {10.1103/PhysRevLett.131.220601},
  urldate = {2025-05-16},
}

@article{gale2020,
  title = {Optimized Fast Gates for Quantum Computing with Trapped Ions},
  author = {Gale, Evan P. G. and Mehdi, Zain and Oberg, Lachlan M. and Ratcliffe, Alexander K. and Haine, Simon A. and Hope, Joseph J.},
  year = {2020},
  month = may,
  journal = {Phys. Rev. A},
  volume = {101},
  number = {5},
  pages = {052328},
  publisher = {Amer Physical Soc},
  address = {College Pk},
  issn = {2469-9926, 2469-9934},
  doi = {10.1103/PhysRevA.101.052328},
  urldate = {2025-05-16}
}

@article{woods2024,
  title = {Coupling Conduction-Band Valleys in {{SiGe}} Heterostructures via Shear Strain and {{Ge}} Concentration Oscillations},
  author = {Woods, Benjamin D. and Soomro, Hudaiba and Joseph, E. S. and Frink, Collin C. D. and Joynt, Robert and Eriksson, M. A. and Friesen, Mark},
  year = {2024},
  month = may,
  journal = {npj Quantum Information},
  volume = {10},
  number = {1},
  pages = {54},
  publisher = {Nature Publishing Group},
  issn = {2056-6387},
  doi = {10.1038/s41534-024-00853-6},
  urldate = {2025-07-25},
  copyright = {2024 The Author(s)}
}

@article{bedell2014,
  title = {Strain Scaling for {{CMOS}}},
  author = {Bedell, S. W. and Khakifirooz, A. and Sadana, D. K.},
  year = {2014},
  month = feb,
  journal = {MRS Bulletin},
  volume = {39},
  number = {2},
  pages = {131--137},
  issn = {0883-7694, 1938-1425},
  doi = {10.1557/mrs.2014.5},
  urldate = {2025-07-25}
}

@article{kjaergaard2020,
  title = {Superconducting {{Qubits}}: {{Current State}} of {{Play}}},
  shorttitle = {Superconducting {{Qubits}}},
  author = {Kjaergaard, Morten and Schwartz, Mollie E. and Braum{\"u}ller, Jochen and Krantz, Philip and Wang, Joel I.-J. and Gustavsson, Simon and Oliver, William D.},
  year = {2020},
  month = mar,
  journal = {Annual Review of Condensed Matter Physics},
  volume = {11},
  number = {Volume 11, 2020},
  pages = {369--395},
  publisher = {Annual Reviews},
  issn = {1947-5454, 1947-5462},
  doi = {10.1146/annurev-conmatphys-031119-050605},
  urldate = {2025-07-30}
}

@article{sawano2008,
  title = {Introduction of {{Uniaxial Strain}} into {{Si}}/{{Ge Heterostructures}} by {{Selective Ion Implantation}}},
  author = {Sawano, Kentarou and Hoshi, Yusuke and Yamada, Atsunori and Hiraoka, Yoshiyasu and Usami, Noritaka and Arimoto, Keisuke and Nakagawa, Kiyokazu and Shiraki, Yasuhiro},
  year = {2008},
  month = dec,
  journal = {Applied Physics Express},
  volume = {1},
  pages = {121401},
  issn = {1882-0778, 1882-0786},
  doi = {10.1143/APEX.1.121401},
  urldate = {2025-08-21}
}

@article{forn-diaz2019,
	title = {Ultrastrong coupling regimes of light-matter interaction},
	volume = {91},
	doi = {10.1103/RevModPhys.91.025005},
	number = {2},
	journal = {Reviews of Modern Physics},
	author = {Forn-Díaz, P.},
	year = {2019},
}

@article{niemczyk2010,
  title = {Circuit Quantum Electrodynamics in the Ultrastrong-Coupling Regime},
  author = {Niemczyk, T. and Deppe, F. and Huebl, H. and Menzel, E. P. and Hocke, F. and Schwarz, M. J. and {Garcia-Ripoll}, J. J. and Zueco, D. and H{\"u}mmer, T. and Solano, E. and Marx, A. and Gross, R.},
  year = {2010},
  month = oct,
  journal = {Nature Physics},
  volume = {6},
  number = {10},
  pages = {772--776},
  publisher = {Nature Publishing Group},
  issn = {1745-2481},
  doi = {10.1038/nphys1730},
  urldate = {2025-09-28},
  copyright = {2010 Springer Nature Limited}
}

@misc{ahn2024,
  title = {Single-Qubit Quantum Gate at an Arbitrary Speed},
  author = {Ahn, Seongjin and Park, Kichan and Cho, Daehee and Lim, Mikyoung and Choi, Taeyoung and Moskalenko, Andrey S.},
  year = 2024,
  month = dec,
  number = {arXiv:2412.19561},
  eprint = {2412.19561},
  primaryclass = {quant-ph},
  publisher = {arXiv},
  doi = {10.48550/arXiv.2412.19561},
  urldate = {2025-09-28},
  archiveprefix = {arXiv}
}

@article{scheuer2014,
  title = {Precise Qubit Control beyond the Rotating Wave Approximation},
  author = {Scheuer, Jochen and Kong, Xi and Said, Ressa S and Chen, Jeson and Kurz, Andrea and Marseglia, Luca and Du, Jiangfeng and Hemmer, Philip R and Montangero, Simone and Calarco, Tommaso and Naydenov, Boris and Jelezko, Fedor},
  year = {2014},
  month = sep,
  journal = {New Journal of Physics},
  volume = {16},
  number = {9},
  pages = {093022},
  publisher = {IOP Publishing},
  issn = {1367-2630},
  doi = {10.1088/1367-2630/16/9/093022},
  urldate = {2025-09-28}
}

@article{yudilevich2023,
  title = {Coherent Manipulation of Nuclear Spins in the Strong Driving Regime},
  author = {Yudilevich, Dan and Salhov, Alon and Schaefer, Ido and Herb, Konstantin and Retzker, Alex and Finkler, Amit},
  year = {2023},
  month = nov,
  journal = {New Journal of Physics},
  volume = {25},
  number = {11},
  pages = {113042},
  publisher = {IOP Publishing},
  issn = {1367-2630},
  doi = {10.1088/1367-2630/ad0c0b},
  urldate = {2025-09-28}
}

@article{costa2025,
  title = {Buried {{Unstrained Germanium Channels}}: {{A Lattice}}-{{Matched Platform}} for {{Quantum Technology}}},
  shorttitle = {Buried {{Unstrained Germanium Channels}}},
  author = {Costa, Davide and Del Vecchio, Patrick and Hudson, Karina and Stehouwer, Lucas E. A. and Tosato, Alberto and Degli Esposti, Davide and Calvi, Vladimir and Moreschini, Luca and Lodari, Mario and Bosco, Stefano and Scappucci, Giordano},
  year = 2026,
  month = may,
  journal = {Advanced Science},
  pages = {e00066},
  issn = {2198-3844, 2198-3844},
  doi = {10.1002/advs.202600066},
  urldate = {2026-06-24}
}

@article{mauro2025,
  title = {Hole Spin Qubits in Unstrained {{Germanium}} Layers},
  author = {Mauro, Lorenzo and Rodr{\'i}guez, Mauricio J. and {Rodr{\'i}guez-Mena}, Esteban A. and Niquet, Yann-Michel},
  year = 2025,
  month = oct,
  journal = {npj Quantum Information},
  volume = {11},
  number = {1},
  pages = {167},
  publisher = {Nature Publishing Group},
  issn = {2056-6387},
  doi = {10.1038/s41534-025-01108-8},
  urldate = {2025-11-07},
  copyright = {2025 The Author(s)}
}

@article{fuchs2009,
  title = {Gigahertz {{Dynamics}} of a {{Strongly Driven Single Quantum Spin}}},
  author = {Fuchs, G. D. and Dobrovitski, V. V. and Toyli, D. M. and Heremans, F. J. and Awschalom, D. D.},
  year = 2009,
  month = dec,
  journal = {Science},
  volume = {326},
  number = {5959},
  pages = {1520--1522},
  issn = {0036-8075, 1095-9203},
  doi = {10.1126/science.1181193},
  urldate = {2026-03-25}
}

@article{mauro2025a,
  title = {Strain Engineering in {{Ge}}/{{Ge}}-{{Si}} Spin-Qubit Heterostructures},
  author = {Mauro, Lorenzo and {Rodr{\'i}guez-Mena}, Esteban A. and Martinez, Biel and Niquet, Yann-Michel},
  year = 2025,
  month = feb,
  journal = {Phys. Rev. Applied},
  volume = 23,
  number = 2,
  pages = 024057,
  issn = {2331-7019},
  doi = {10.1103/PhysRevApplied.23.024057},
  urldate = {2026-04-16}
}

@article{sawano2016,
  title = {({{Invited}}) {{Anisotropic Strain Introduction}} into {{Si}}/{{Ge Hetero Structures}}},
  author = {Sawano, Kentarou and Konoshima, Shiori and Yamanaka, Junji and Arimoto, Keisuke and Nakagawa, Kiyokazu},
  year = 2016,
  month = aug,
  journal = {ECS Trans.},
  volume = {75},
  number = {8},
  pages = {563--569},
  issn = {1938-5862, 1938-6737},
  doi = {10.1149/07508.0563ecst},
  urldate = {2026-03-26}
}

@article{takeuchi2011,
  title = {{{Ge}}$_{1-x}${{Sn}}$_{x}$ Stressors for Strained-{{Ge CMOS}}},
  author = {Takeuchi, S. and Shimura, Y. and Nishimura, T. and Vincent, B. and Eneman, G. and Clarysse, T. and Demeulemeester, J. and Vantomme, A. and Dekoster, J. and Caymax, M. and Loo, R. and Sakai, A. and Nakatsuka, O. and Zaima, S.},
  year = 2011,
  month = jun,
  journal = {Solid-State Electronics},
  volume = 60,
  number = 1,
  pages = {53--57},
  issn = 00381101,
  doi = {10.1016/j.sse.2011.01.022},
  urldate = {2026-04-01}
}

@article{casse2009,
  title = {A Comprehensive Study of Magnetoresistance Mobility in Short Channel Transistors: {{Application}} to Strained and Unstrained Silicon-on-Insulator Field-Effect Transistors},
  shorttitle = {A Comprehensive Study of Magnetoresistance Mobility in Short Channel Transistors},
  author = {Cass{\'e}, M. and Rochette, F. and Thevenod, L. and Bhouri, N. and Andrieu, F. and Reimbold, G. and Boulanger, F. and Mouis, M. and Ghibaudo, G. and Maude, D. K.},
  year = 2009,
  month = apr,
  journal = {J. Appl. Phys.},
  volume = {105},
  number = {8},
  pages = {084503},
  issn = {0021-8979, 1089-7550},
  doi = {10.1063/1.3097764},
  urldate = {2026-04-01}
}
\bibliographystyle{naturemag}

{\bf Methods}

Our calculations on the Ge$_{120}$/(Si$_{0.2}$Ge$_{0.8}$)$_{60}$ QW are performed using a $5\times5$ supercell with 120 Ge and 60 SiGe atomic layers along the [001] growth direction. We employ the VFF~\cite{Peter2005,Zunger_JAP1998} approach to optimize the lattice structure by minimizing the strain energy, which has been previously applied to conventional semiconductors, including Si and Ge~\cite{WangLW_PRB2000,Luo_NC2013,Luo2017,Xiong_PRB2021,Wang1995,Luo2010}. We perform the atomistic SEPM calculations based on the optimized structures. The electronic structures of Ge/SiGe QWs are obtained by directly diagonalizing the band Hamiltonian, utilizing the folded spectrum method in a plane-wave basis~\cite{Wang_JCP1994}. An energy cutoff of 8.2 Ry is used to select the plane-wave basis. A $16\times16\times16$ grid in real space is used for each eight-atom cubic cell. To estimate the $k$-linear SOC parameters, we fit the spin-splitting of the HH1 along the different routes upon application of a vertical electric field.
\yxadd{The effect of strain caused by thermal mismatch is ignored due to their small magnitude compared to the uniaxial strain applied~\cite{Abadillo-Uriel2023}.}

\section*{}
\paragraph*{Acknowledgements}
J.-W.~L. was funded by the Natural Science Foundation of China (NSFC) under grant No. 12525402 and the CAS Project for Young Scientists in Basic Research under grant No. YSBR-026.
S.~G. was funded by the NSFC under grant No. 12374078.
Y.~L. was funded by the NSFC under grant No. 12504094, and the Postdoctoral Fellowship Program of China Postdoctoral Science Foundation under grant No. GZC20252226.
\paragraph*{Author contributions}
J.-W.~L. conceived the research concept and led the writing of the manuscript. S.~G. and J.-W.~L. supervised the research. Y.-X.~W., Y.~L. and S.~G. carried out the first-principles calculation, data analysis, and analytical modeling. All authors contributed to data analysis, scientific discussion, and review of the manuscript.

\paragraph*{Competing interests}
There are no competing interests to declare.

\paragraph*{Correspondence}
Correspondence should be addressed to Shan Guan or Jun-Wei Luo.

\paragraph*{Additional Information}
Supplementary Information is available in the online version of the paper.

\end{document}